# Hot phonon effects on high-field transport in 2DEG GaN.

Lee Smith, Peter Brookes Chambers, Daniel R. Naylor & Angela Dyson

*Abstract*—The effects of confinement on electron transport in GaN have been studied via an ensemble Monte-Carlo code. Excellent agreement is obtained with experimental data from the literature up to moderate fields. In agreement with experimental results, negative-differential-conductivity is not observed in velocity-field curves. The reasons for this are discussed in detail. The dynamics of the non-equilibrium confined electron-LO phonon system is studied via a bulk phonon spectrum. In contrast to the bulk electron case hot or non-equilibrium phonons do not play a significant role in determining the transport properties. This is explained via the energy and momentum conservation rules for the polar optical phonon scattering. In addition, impact ionization is shown to be insignificant for the applied fields considered.

## I. INTRODUCTION

There have been very few reports on the inclusion of electron confinement into Monte-Carlo (MC) semiconductor transport codes for III-V semiconductors [1] and even fewer on the inclusion of non-equilibrium phonons into those codes [2,3]. In GaN devices such as HEMTs agreement between bulk MC results and experimental data is very poor suggesting that electron confinement should be considered. In GaN devices self-heating has long been at least partially attributed to the non-equilibrium phonon effect. The longitudinal optical phonon energy is very large and it is possible to drive the phonon population far from equilibrium. In bulk the hot-phonon effect limits the saturated velocity of the electrons [4, 5]. Hot phonons were included in the work of Ramonas *et. al.* [3] though applied electric fields in excess of 35 kV/cm were not considered. It has been proposed that hot electrons may be responsible for the degradation of AlGaN/GaN & InAlN/GaN HFET's [6] with electron temperatures measured experimentally at 2600 K and 2850 K respectively [7,8]. While such high temperatures may be highly localized it is clear that hot phonon effects cannot be neglected. Sophisticated MC codes are usually coupled to Poisson solvers [9] and often isothermal or electrothermal solvers[10,11]. Data from these tends to be published in the form of current voltage characteristics which are necessarily steady-state. This data is of little use in diagnosing the role of hot-electrons in device degradation. In general the mobilities observed in GaN heterostructures are highly device dependent with major differences noted for Ga-polar or N-polar structures [12].

In this communication, we study the steady state transport in a GaN 2DEG with a dynamic phonon distribution which is updated periodically.

## II. METHOD

We have developed an ensemble 2DEG MC simulation in which hot phonons are included using the algorithm outlined by Jacaboni & Lugli [13]. The scattering and phonon occupation tables are updated every 25 fs. Details of the implementation for bulk semiconductors with non-equilibrium phonons can be found elsewhere [5].  We adopt the commonly used approximation that GaN has a zinc blende structure. This assumption reduces the number of the phonon modes. The 2DEG is assumed to form at the interface with an AlGaN barrier giving electron confinement in the z-direction. The first 2 sub-bands of the resulting triangular well are considered making use of Fang-Howard wavefunctions. In 2D the scattering rates are in general calculated numerically; mechanisms included in the simulation are: non-polar optical, polar optical phonon, acoustic deformation potential, alloy and piezoelectric. Scattering may take place within a sub-band – intra sub-band scattering or between sub-bands – inter sub-band scattering. The zinc blende approximation allows us to consider a single LO mode and we ignore the dispersion since it is small. Screening and degeneracy effects are not currently included. The simulations are performed with 100,000 superparticles. The main parameters [5] used in the model are given in table 1. The phonon lifetime is known to be a function of electron density and temperature in GaN [14] with measured lifetimes between 2.5 ps and around 0.1 ps [15]. A phonon lifetime of 1 ps was chosen and the lattice temperature is 300 K throughout [3].

## III. RESULTS AND DISCUSSION

### A. 2DEG with equilibrium phonons

The results from the code are compared with experimental data and bulk MC results. Figure 1 shows the velocity field characteristics generated by our code with phonons at the lattice

Manuscript received  ??, 2024. This work was supported by the Office of Naval Research

A. Dyson is with School of Maths, Stats & Physics, Newcastle University, UK ( e-mail angela.dyson@ncl.ac.uk)
L Smith was with University of Hull, Hull, HU6 7RX, UK.

TABLE I
MATERIAL PARAMETERS

|  | GaN |
|---|---|
| Effective mass Γ valley | 0.2 |
| Effective mass upper valley | 1.0 |
| Non-polar optical phonon energy (meV) | 91.2 |
| Polar optical phonon energy (meV) | 91.2 |
| Static dielectric constant | 8.9 |
| High frequency dielectric constant | 5.35 |
| Longitudinal sound velocity ($10^3$ m/s) | 6.56 |
| Transverse sound velocity ($10^3$ m/s) | 2.68 |
| Acoustic deformation potential (eV) | 8.3 |
| Optical Phonon lifetime (ps) | 1 |

temperature. Here the bulk simulation uses an intervalley separation of 1.9 eV as is typical in the
literature. The resulting bulk velocities are very high and so far not observed in bulk experiments. The 2DEG simulation employs a confining field that would give an effective well width of 4.1 nm. With sub-band energies of 0.290 eV and 0.507 eV. From the data a value of 2600 cm$^2$/ V S is obtained for the mobility which compares with an experimentally determined value of 1670 cm$^2$/ V S [16].

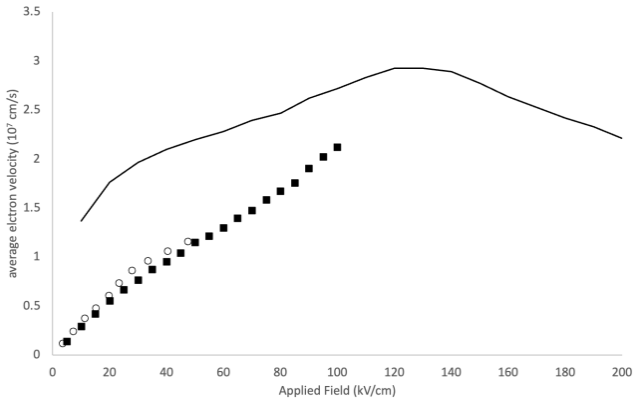

Fig. 1. Steady-state velocity-field curve for bulk GaN solid line, filled squares 4.1 nm 2DEG (this work) and data from [16] open circles.

The 2DEG results show excellent agreement with the experimental data of Palacios *et al.* [16].
Figure 2 shows the velocity field characteristics as a function of effective well width. A narrower width has a higher sheet density. The velocity shows a linear dependence on effective well width. We conclude that a bulk implementation is of limited use when studying GaN 2DEG's.

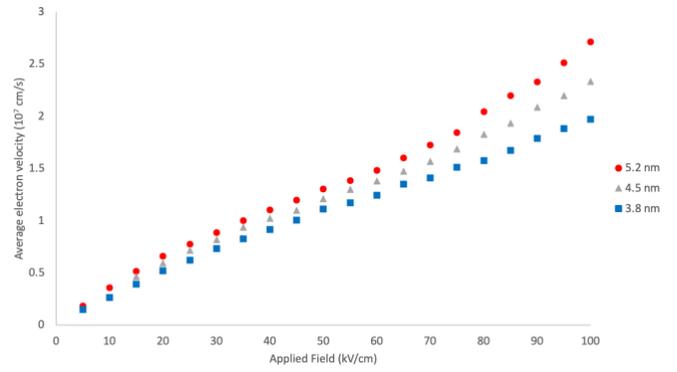

Fig. 2. Steady-state velocity-field curve for different effective well widths. Circles 5.2 nm, triangles 4.5 nm, squares 3.8 nm.

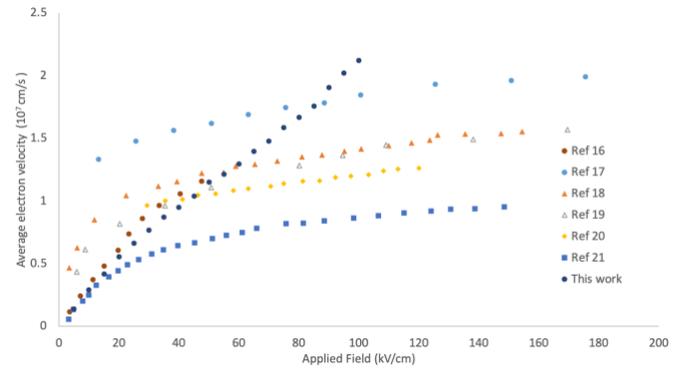

Fig. 3. Experimental results of electron velocity versus applied field.

In Figure 3 experimental velocity data from AlGaN/GaN or AlGaN/AlN/GaN heterostructures are presented along with the results of this work. The spread in the experimental results is considerable. While the velocity does show a flattening off at higher fields in the experimental data there is no evidence of a negative-differential-conductivity (NDC). Atmaca *et al.* observed an NDC in an AlGaN/AlN/GaN heterostructure beyond 200 kV/cm [17], our results would suggest that any NDC originates in the bulk not the 2DEG. In bulk the NDC derives from the transfer of electrons from the lower gamma valley to one of the satellite valleys where the effective mass is higher. Conservation of energy dictates that the velocity must reduce. In a 2DEG the same mechanism could be in play either due to electrons occupying continuum i.e. bulk states or if the well were wide enough to provide sufficient subbands to allow a transition from the gamma valley to one of the higher valleys. As the effective well width in GaN devices is predominantly below 5nm this mechanism is effectively eliminated as the 2 existing sub bands are in the gamma valley.
The pulse durations of the I-V measurements are 20 ns at 77K [17] 1 ns [18], 2 ns [19], 80 ns [20] and 200 ns [16,21] all at room temperature. Its clear that the peak velocity reached depends on the pulse duration of the I-V measurements; with shorter pulse durations leading to higher velocities. This would suggest that self-heating may be contributing to the velocity saturation in the experiments.

Self-heating is known to be an issue in GaN. The large polar-optical phonon energy and rapid emission rate mean that the phonon population can be quickly driven out of equilibrium. In this case the electron velocity can be limited by non-equilibrium phonons and would be dependent on the phonon lifetime.

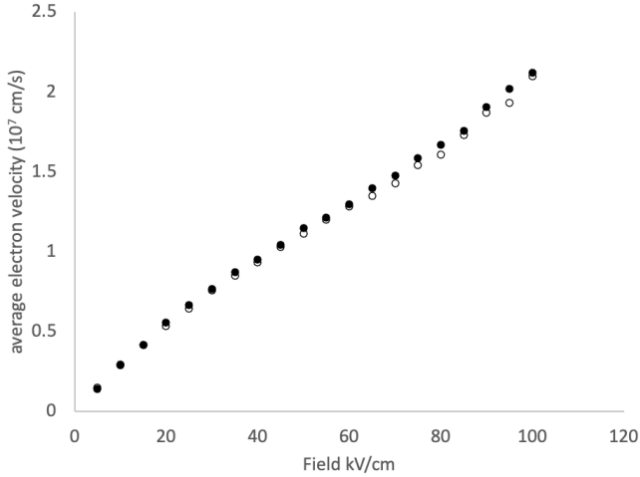

Fig. 4. Steady-state 2DEG velocity-field curve with non-equilibrium phonons open circles and closed circles equilibrium phonons.

### B. 2DEG with non-equilibrium phonons

Figure 4 shows the velocity field curves with non-equilibrium phonons and with phonons at the lattice temperature. There is very little difference between the two. This is in contrast to the bulk case where non-equilibrium phonons play a significant role in the transport properties [5]. In the bulk case peak electron velocities are 12% lower when non-equilibrium phonons are considered. In the bulk, non-equilibrium phonons lead to diffusive heating, increasing electron energies. While the "hot phonon effect" is present in 2D it is much smaller than in the bulk case.

The reason for this disparity between bulk and 2DEG is that in order to satisfy energy and momentum conservation the confined electron has access to a smaller region of phonon momentum space than is the case in bulk. That is phonons are emitted in a small volume of q-space. As a result, the phonon occupancy does not evolve away from equilibrium across the spectrum of wavevectors as it does in the bulk. Figure 5 shows the phonon occupancy as a function of wavevector for both 2D and bulk simulations. Note that the scale is 10 times larger for the bulk transverse phonon wavevector. There is a preference for forward scattering, this is well known for optical phonons. At the lattice temperature the phonon occupation, n(w) takes a value of 0.03 for GaN. Here a peak of around 2.5 to 3 is reached, which would correspond to a phonon temperature of 3200 K to 3700 K, this compares favorably with experimental results with an electron temperature of 2850 K [8]. Phonon temperature is usually taken to be equivalent to the hot electron temperature in the relaxation time approach. [1]

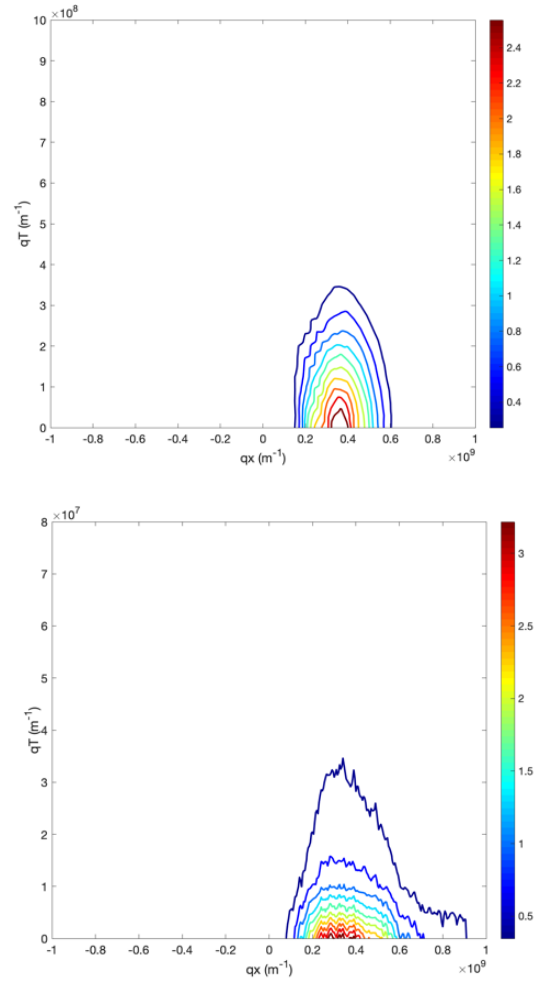

Fig. 5. Non-equilibrium phonon occupancy as a function of wavevector in bulk (top) at an applied field of 50 kV/cm and in 2D (bottom) at an applied field of 70 kV/cm. qx being the phonon wavevector parallel to the applied field while qT is that transverse to the applied field.

It is to be expected that the phonon temperature would be higher in the 2D simulation than in experiment as in the simulation there is no mechanism by which the electrons can escape the 2D channel. In experiment they might escape into the barrier or the substrate. The 2D simulation does not currently include impact ionization. Electron energies of 3.4 eV i.e close to the band gap are required for impact ionization. Our bulk MC simulations show that there are not significant numbers of hot electrons reaching these energies, so impact ionization is not expected to be significant unless the electric fields are much higher than those studied here. Fig 6 shows the distribution of bulk electrons at an applied field of 600 kV/cm with impact ionization included. The numbers of electrons with energies above 3.4 eV is a fraction of a percent. Reigrotzki calculated the impact ionization scattering rate via a MC code with a full band implementation for GaN [22]. Electron energies in excess of 8 eV were required to achieve significant scattering rates i.e. rates similar to polar optical phonon scattering. It therefore seems unlikely that impact ionization plays a significant role. The phonon lifetime is known to vary as a function of sheet density; our choice of 1 ps would be on the high side for very high sheet densities. A smaller phonon lifetime would lessen the hot-phonon effect further.

Taken together these results suggest that there is another mechanism by which electrons dissipate energy in a 2DEG. Brazzini *et al.* have proposed that the electroluminescence observed from HEMT structures derives from Bremsstrahlung radiation emitted by electrons decelerating following a scattering event [8].

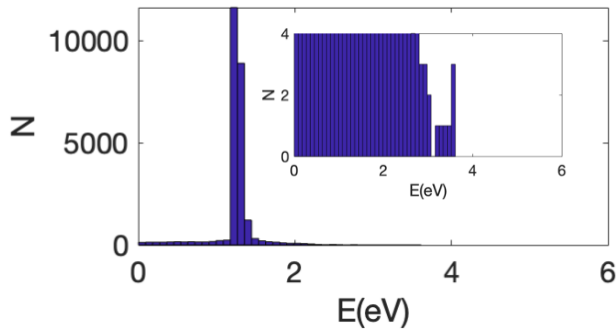

Fig. 6. Electron distribution (number of superparticles as a function of energy) at an applied field of 600 kV/cm from a bulk MC simulation with impact ionization included. Inset shows a zoom in, so high energy population can be seen.

## III. CONCLUSIONS

We have studied the effects of confinement and non-equilibrium phonons on the transport properties of electrons in a GaN 2DEG with 2 subbands. Under steady-state conditions excellent agreement is obtained with experimental results up to moderate applied fields. The non-equilibrium phonons do not play as big a role as is the case in bulk. This is due to the small volume of q-space available during scattering that satisfies energy and momentum conservation. In contrast to the bulk case where non equilibrium phonons drive significant diffusive heating. These results show that the velocity saturation observed in experiments does not originate from the hot phonon effect. It has also been demonstrated through bulk simulation results that impact ionization is not expected to play a significant role. These results lend support recent literature proposing that hot electrons relax to lower energy states via emission of Bremsstrahlung [8] during scattering.

ACKNOWLEDGMENT

This work was supported by the Office of Naval Research sponsored by Dr. Paul Maki grant no. N00014-18-1-2463.

**Lee Smith** received the M.Phys. degree (Honours) in physics, in 2015, and the Ph.D. degree in 2021, both from the University of Hull, U.K. He is now with VISR Dynamics, Hull.

**Peter Brookes Chambers** received the MPhys degree (Honours) in physics from Newcastle University in 2019 and is studying for a PhD. Their research interests are methods for impact ionization in monte-carlo codes.

**Daniel R. Naylor** received the B.Sc. degree (Honours) in physics, in 2008, and the Ph.D. degree in 2012, both from the University of Hull, U.K. In 2013, he joined Tessella. He returned to Hull in 2015 and moved to Newcastle University in 2017. He is now with Opencast.

**Angela Dyson** received the BSc (Honours) degree in theoretical physics and the PhD degree in theoretical plasma physics from the University of Essex, Colchester, UK, in 1994 and 1999 respectively. In 2008 she joined the Department of Physics at the University of Hull; where she was promoted to reader in 2016. She joined Newcastle University in 2017 where she is Director of Physics. She has worked in the area of semiconductor device theory and modeling for almost 25 years. Her current research interests include high field transport, monte carlo device simulation and wide gap and ultra-wide gap semiconductors.